\documentstyle[aps,epsf,twocolumn]{revtex}

\def\be{\begin{equation}}
\def\ee{\end{equation}}
\def\bea{\begin{eqnarray}}
\def\eea{\end{eqnarray}}

\def\t12h{\frac{\theta_{12}}{2}}

\def\eps{\varepsilon}
\def\ep{\varepsilon}
\def\ii{{\rm i}}
\def\stackreb#1#2{\mathrel{\mathop{#2}\limits_{#1}}}
\def\res#1{\stackreb{#1}{\rm Res}}
\def\r#1{(\ref{#1})}
\def\nn{\nonumber\\}
\def\up{\uparrow}
\def\da{\downarrow}

\def\NPB#1#2#3{{ Nucl. Phys.} {\bf B#1}, #2 (#3)}

\def\PRD#1#2#3{{ Phys. Rev.} {\bf D#1}, #2 (#3)}
\def\PRB#1#2#3{{ Phys. Rev.} {\bf B#1}, #2 (#3)}
\def\PRL#1#2#3{{ Phys. Rev. Lett.} {\bf #1}, #2 (#3)}

\begin{document}

\title{\hfill{\rm OUTP-98-10S}\\
{\bf Formfactors in the half-filled Hubbard model}}

\author{Fabian H.L. E\ss ler$^{(a)}$ and Vladimir E. Korepin$^{(b)}$}
\address{$^{(a)}$ Department of Physics, Theoretical Physics,
        Oxford University\\ 1 Keble Road, Oxford OX1 3NP, United Kingdom}
\address{$^{(b)}$ Institute for Theoretical Physics, SUNY at Stony
Brook, Stony Brook, NY 11794-3840, USA}
\maketitle
\begin{abstract}
\par
We consider dynamical spin-spin correlation functions in the one
dimensional repulsive half-filled Hubbard model. We propose an exact
expression for the two spinon formfactor of spin operators. We use
this to derive the two spinon contribution to the dynamical structure
factor.
\end{abstract}

\begin{narrowtext}
\section{Introduction}
The calculation of dynamical correlation functions is one of the main
challenges in low dimensional theories of statistical and condensed
matter physics. The large distance behaviour (corresponding to very
low energies) can be effectively extracted for critical theories by
using bosonization or Bethe Ansatz and conformal field theory
techniques \cite{lowE,vladb,GNT,reprint}. The results obtained in this way
for models like the Heisenberg XXZ chain have been
successfully applied to neutron scattering experiments of quasi-1D
magnetic insulators, see e.g. \cite{magnets,cubenz}. However, from an
experimental point of view not only very low energies are of
interest. In magnetic materials described by the spin-$1/2$ Heisenberg
model \cite{magnets,cubenz} the scattering continuum is measured up to
energies of twice the exchange, which is outside the realm of
reliability of the above methods. It therefore would be useful to have
a method for the calculation of (Fourier tansforms of) correlation
functions at all energy scales. Another setting that is outside the
realm of the above methods are systems with a spectral gap, see
e.g. \cite{cubenz,cugeo3,gaptheory,et}. Some systems with spectral gap
can be dealt with by the {\sl formfactor} approach to quantum
correlation functions \cite{karowski,smirnov,luk,fff,et}. The advantage of
this method is that it is in principle not constrained to very low
energies. The formfactor approach can also be implemented for critical
systems, in particular this has been done for the Heisenberg XXX chain
\cite{miwa,bou}. The results are useful for the analysis of the 
observed two-spinon scattering continuum in magnetic insulators like
${\rm KCuF_3}$. In the present paper we consider the problem of calculating 
formfactors in the one dimensional half-filled Hubbard model. In a way
this model is the ``parent'' of the Heisenberg model used for the
description of the aforementioned magnetic insulators: 
the charge sector is gapped and the spin sector for large repulsion between
electrons is described by an effective Heisenberg Hamiltonian. However,
if the system is probed at energy scales of the order of the Coulomb 
repulsion e.g. by photoemission experiments, which measure 
the single-particle spectral function, the itinerant nature of the system
emerges and it is necessary to calculate correlation functions in the
half-filled Hubbard chain in order to describe the experiments.
In the present context it is of particular interest to calculate formfactors
of the electronic creation and annihilation operators, as these results
could be directly applied to the ARPES data of \cite{srcuo2}. The easier 
problem of determining formfactors of the spin operators in the Hubbard 
model is the subject of the present work.

We consider the repulsive half filled Hubbard model \cite{lw}. 
The Hamiltonian is ($U>0$)
\bea
H(U) &=& -\sum_{j=1}^L\sum_{\sigma=\up ,\da} 
\left(c^\dagger_{j,\sigma} c_{j+1,\sigma} + c^\dagger_{j+1,\sigma}
c_{j,\sigma}\right)\nn
&&+ 4U\sum_{j=1}^L (n_{j,\up}-{1\over 2})(n_{j,\da}-{1\over 2})\ .
\label{hamil}
\eea
This Hamiltonian exhibits an $SO(4)$ symmetry \cite{yang},
i.e. it commutes with the generators
\bea
S &=& \sum_{j=1}^L c^\dagger_{j,\up} c_{j,\da}\ , \
S^z = \sum_{j=1}^L {1\over 2} (n_{j,\da}-n_{j,\up}) \ ,\label{spin}\\
\eta &=& \sum_{j=1}^L (-1)^jc_{j,\up} c_{j,\da}\ , \
\eta^z = {1\over 2}\sum_{j=1}^L (n_{j,\up}+n_{j,\da}  - 1)\ .
\label{eta}
\eea

The complete spectrum of low-lying excitations was determined in
\cite{smat}. It consists of pairs of scattering states of four elementary
excitations, which form the fundamental representation of $SU(2)\times
SU(2)$. There is a doublet of uncharged, gapless spin-1/2 particles
called spinons. Their energy and momentum (as functions of the
rapidity variable $\beta$) are given by \cite{woy,smat}
\bea
p_s(\beta) &=&
\frac{\pi}{2}-\int_0^\infty\frac{d\omega}{\omega}\frac{J_0(\omega)\
\sin(\omega 2U\beta/\pi)}{\cosh(\omega U)}\ ,\nn
\eps_s(\beta) &=&
2\int_0^\infty\frac{d\omega}{\omega}\frac{J_1(\omega)\
\cos(\omega 2U\beta/\pi)}{\cosh(\omega U)}\ ,
\eea
where $J_{0,1}$ are Bessel functions.
The other two elementary excitations carry charge $\pm
e$ but no spin. They are called holon and antiholon and have a gap
proportional to $U$. Their energy and momentum are
\bea
\eps_c(k) &=& 2U+2\ {\rm cos}k + 2 \int_{0}^\infty {d\omega\over \omega}
{J_1(\omega) {\rm cos}(\omega\ {\rm sin}k)e^{-\omega U}\over {\rm
cosh}(\omega U)}\nn
p_c(k) &=& \pi/2-k - \int_{0}^\infty {d\omega\over \omega} {J_0(\omega) {\rm
sin}(\omega\ \sin k)e^{-\omega U}\over {\rm cosh}(\omega U)}.
\eea

\section{Dynamical structure factor}

In this paper we consider the dynamical structure factor, which is the
Fourier transform of the dynamical spin-spin correlation function and
which is measured by inelastic neutron scattering. The formfactor
expansion is given by 
\bea
&&S(\omega,p)=\int_{-\infty}^\infty dt\ \sum_{m=-\infty}^\infty
e^{i\omega t + i p m}\ \langle 0|\sigma^+_m(t)\sigma^-_0(0)|0\rangle\nn
&&=\sum_{n=2}^\infty\frac{1}{n!}
\int_{-\infty}^\infty dt\ \sum_{m=-\infty}^\infty e^{i\omega t + i p m}
\prod_{k=1}^n\left(\int_{-\infty}^\infty \frac{d\beta_k}{2\pi}\right)\nn
&&\times\langle 0|\sigma^+_m(t)|\beta_n\ldots\beta_1\rangle_{\eps_n\ldots\eps_1}\
{_{\eps_1\ldots\eps_n}\langle \beta_1\ldots\beta_n|\sigma^-_0(0)|0\rangle}\nn
&&=\sum_{n=2}^\infty S_n(\omega,p)\ ,
\label{dcf1}
\eea
where the labels $\eps_j$ enumerate the four possible elementary
excitations. Our notation is the following: $\eps=\pm$ denotes a
spinon with spin up/down, $\eps=1,-1$ denotes an antiholon/holon. As
excitations involving holons and antiholons have gaps the main
contribution to the correlation function \r{dcf1} comes from
multi--spinon excitations. By virtue of related results obtained in
other gapless models \cite{muss} we expect that the main contribution
is due to excitations involving only two spinons. Because of the
spin-$SU(2)$ symmetry the 2-spinon contribution is of the form 

\bea
&&S_2(\omega,p)=2\pi\sum_{m=-\infty}^\infty\int_{-\infty}^\infty
\frac{d\beta_1}{2\pi}\int_{-\infty}^\infty \frac{d\beta_2}{2\pi}\nn
&&\times \exp\left(im[p+{p_s}(\beta_1)+{p_s}(\beta_2)]\right)\nn
&&\times\delta(\omega-{\eps_s}(\beta_1)-{\eps_s}(\beta_2))\ \left|\langle 0|
\sigma_1^+|\beta_2 \beta_1\rangle_{--}\right|^2  .
\label{dcf2}
\eea
We propose the following expression for the two-spinon formfactor
\bea
&&f(\beta_1,\beta_2)_{--}=\langle 0|\sigma_1^+|\beta_2
\beta_1\rangle_{--}=\nn
&&c\
A_-(\beta_2-\beta_1)/\left[\sinh(i\pi/4-\beta_1/2) 
\sinh(i\pi/4-\beta_2/2)\right]\ ,\nn
&&A_-(\beta) = \exp\left(-\int_0^\infty \frac{dt}{t}
\frac{\sinh^2(t[1-\beta/i\pi])\exp(t)}{\sinh(2t)\ \cosh(t)}\right) .
\label{a-}
\eea

Here $c$ is the usual common constant factor in all formfactors. 
We presently cannot determine its exact numerical value although it is
possible to obtain an estimate by considering various sum rules.
Let us now provide some evidence for the validity of \r{a-}. 

For integrable relativistic quantum field theories one generically has
a formfactor expansion of the form \r{dcf1}, \r{dcf2} for correlation
functions of local operators. The formfactors themselves are
determined by the following set of axioms \cite{karowski,smirnov}

\noindent
{\bf Axiom 1.} The form factors have the symmetry property
\bea\label{ax1}
&f(\ldots,\beta_i,\beta_{i+1},\ldots)
    _{\ep_1,\ldots,\ep_i,\ep_{i+1},\ldots,\ep_n}
S_{\ep_i,\ep_{i+1}}^{\ep'_i,\ep'_{i+1}} (\beta_i-\beta_{i+1})\nn
&\quad=f(\ldots,\beta_{i+1},\beta_{i},\ldots)
_{\ep_1,\ldots,\ep'_{i+1},\ep'_{i},\ldots,\ep_n}
\eea

\noindent
{\bf Axiom 2.} The formfactors fulfil the
tensor-valued$\ \ \ $ Riemann-Hilbert problem
\be
f(\beta_1\ldots\beta_n+2\pi\ii)_{\ep_1\ldots\ep_n}=
f(\beta_n\beta_1\ldots\beta_{n-1})_{\ep_n\ep_1,\ldots,\ep_{n-1}}
\label{ax2}
\ee

\noindent
{\bf Axiom 3.} In the absence of bound states the only
singularities of $f(\beta_1,\ldots,\beta_n)_{\ep_1,\ldots,\ep_n}$ for
$n\geq 3$ are at the points $\beta_i=\beta_j+ \pi\ii$, $i>j$. These
singularities are first order poles (annihilation poles) with residues 
\bea\label{ax3}
&&\ii\res{\beta_n=\beta_{n-1}+\pi\ii}
f(\beta_1\ldots\beta_n)_{\ep_1,\ldots,\ep_n}=\nn
&&f(\beta_1,\ldots,\beta_{n-2})
_{\ep'_1,\ldots,\ep'_{n-2}}\delta_{\ep_n,-\ep_{n-1}}\biggl(
\delta_{\ep_1}^{\ep'_1}\ldots \delta_{\ep_{n-2}}^{\ep'_{n-2}}-\nn
&& S_{\tau_1,\ep_{1}}^{\ep'_{n-1},\ep'_{1}}
(\beta_{n-1}-\beta_1)
\ldots
S_{\ep_{n-1},\ep_{n-2}}^{\tau_{n-3},\ep'_{n-2}}(\beta_{n-1}-\beta_{n-2})
   \biggr).
\eea
Here $S^{\alpha\beta}_{\gamma\delta}(\beta)$ is the 2-particle
scattering matrix of the theory under consideration. Although these
axioms are based on crossing symmetry and relativistic energy-momentum
relations, there is evidence that the axioms hold true even for
formfactors of certain nonrelativistc lattice models. In particular,
it was shown by Pakuliak \cite{paku} that the known formfactors
\cite{miwa} of the XXZ Heisenberg lattice model fulfill Axioms 1-3. We
believe that this will be true for integrable models of statistical
mechanics and condensed matter physics as long as the ground state is
a singlet under the action of an infinite dimensional symmetry algebra
like a Yangian or $U_q(\widehat{sl(2)})$. In particular, the
formfactor \r{a-} can be seen to fulfil Axioms 1-3
\bea
f(\beta_1,\beta_2+2\pi i)_{--} &=& f(\beta_2,\beta_1)_{--} \ ,\nn
f(\beta_1,\beta_2)_{--} &=& S_{--}(\beta_2-\beta_1)\
f(\beta_2,\beta_1)_{--} \ ,
\label{12}
\eea
where $S_{--}(\beta)$ is the dressed S-matrix for scattering of
spinons in the spin triplet state with $S^z=-1$
\be
S_{--}(\beta) = \frac{\Gamma(-\beta/2\pi i)}{\Gamma(\beta/2\pi i)}
\frac{\Gamma(1/2+\beta/2\pi i)}{\Gamma(1/2-\beta/2\pi i)}\ .
\ee
We think that the fact that \r{a-} fulfils \r{12} is a good
indication for the correctness of the ``minimal formfactor''
$A_-(\beta)$. The full result \r{a-} can be checked exactly in the
limit $U\to\infty$, where the half-filled Hubbard model \r{hamil} 
reduces to the isotropic spin-1/2 Heisenberg chain 
\be
H_{XXX}= J\sum_{j=1}^L \vec{S}_j\cdot\vec{S}_{j+1}\ ,
\label{heis}
\ee
where the exchange is given by $J=1/U$. The spinon formfactors for
this model have been calculated in \cite{miwa,bou}. The dynamical
structure factor can be represented in the form \r{dcf1}, where the
sum is over all spinon states. The 2-spinon contribution is
\cite{bou} 
\bea
&&S_2^{XXX}(\omega,p)=2\pi\sum_{m=-\infty}^\infty\int_{-\infty}^\infty
\frac{d\beta_1}{2\pi}\int_{-\infty}^\infty \frac{d\beta_2}{2\pi}\nn
&&\times\exp\left(im[p+{p}(\beta_1)+{p}(\beta_2)]\right)\nn
&&\times\delta(\omega-{\eps}(\beta_1)-{\eps}(\beta_2))\ \left|\langle 0|
\sigma_1^+|\beta_2 \beta_1\rangle_{--}\right|^2 
\eea
where
\bea
&&\langle 0|\sigma_1^+|\beta_2 \beta_1\rangle_{--} =c^\prime\
A_-(\beta_2-\beta_1)\nn
&&\hskip1cm\times
\left(\sinh(i\pi/4-\beta_1/2)\sinh(i\pi/4-\beta_2/2)\right)^{-1}
\label{a-xxx}
\eea
Here the dressed energy ${\eps}(\beta)$ and momentum
${p}(\beta)$ are given by
\bea
{\eps}(\beta)=J\pi/2\cosh\beta\ ,\quad {p}(\beta) = {\rm
arccot}(\sinh\beta)\ .
\eea
It is a straightforward to see that in the limit $U\to\infty$
$S_2(\omega,p)$ indeed reduces to $S_2^{XXX}(\omega,p)$. The $U\to
\infty$ limit of $\eps_s(\beta)$ and $p_s(\beta)$ is derived for
example in \cite{ezer}.
We believe that \r{a-} is correct for any value of $U>0$. In
\cite{paku2} an independent derivation for formfactors
of $\sigma^+$ in the isotropic Heisenberg XXX chain \r{heis} was given
by using representation theory of a central extension of a double of
the $sl(2)$-Yangian. For this purpose $\sigma^+$ can be represented as the
density of one of the zeroth level Yangian generators ($E_0$ in the
notation of \cite{yangian}). Because the spin part of the Yangian
\cite{yangian} of the Hubbard model is the same as in the XXX case, we
believe that the XXX result can be ``lifted'' to the half-filled
Hubbard model in the way presented above.

In order to perform the integrals over $\beta_{1,2}$ in \r{dcf2}
we need to discuss some properties of the continuum of two-spinon
excited states. Energy and momentum are given by
\bea
E(\beta_1,\beta_2)&=&\eps_s(\beta_1)+\eps_s(\beta_2)\ ,\nn
P(\beta_1,\beta_2)&=&p_s(\beta_1)+p_s(\beta_2)\ {\rm mod}\ 2\pi ,
\eea
where $\beta_{1,2}\in(-\infty,\infty)$. The upper boundary is obtained
by taking $\beta_1=\beta_2$ and the lower one by taking
$\beta_1=\pm\infty$. We denote the respective dispersion relations
by $\omega_{U,L}(p)$. In Fig.1 we plot the 2-spinon continuum for
$U=5$.

\begin{figure}[ht]
\begin{center}
\noindent
\epsfxsize=0.45\textwidth
\epsfbox{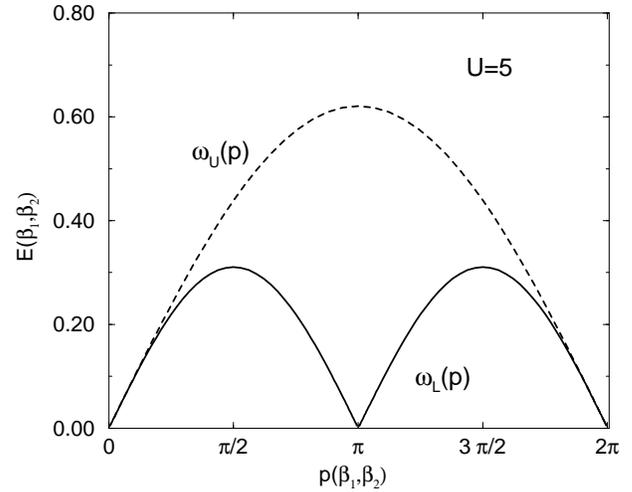}
\end{center}
\caption{\label{fig:disp}%
Continuum of two spinon excitations fur $U=5$
.} 
\end{figure}
In the first Brillouin zone there is a unique solution
$(\bar{\beta_1},\bar{\beta_2})$ to the set of
equations 
\bea
\omega&=&\eps_s(\beta_1)+\eps_s(\beta_2)\ ,\nn
p&=&p_s(\beta_1)+p_s(\beta_2)\ ,
\eea
as long as $\omega$ is chosen within the interval
$(\omega_L(p),\omega_U(p))$. This allows us to taken the integrals
over $\beta_{1,2}$ in \r{dcf2} with the result
\bea
&&S_2(\omega,p)=\bar{c}\Theta(\omega-\omega_L(p))\
\Theta(\omega_U(p)-\omega) |A_-(\bar{\beta_2}-\bar{\beta_1})|^2\nn
&&\times 
\left(\cosh(\bar{\beta_1}/2)
\cosh(\bar{\beta_2}/2)\right)^{-1}\nn
&&\times \left|
\frac{\partial \eps_s(\bar{\beta}_1)}{\partial\bar{\beta}_1}
\frac{\partial p_s(\bar{\beta}_2)}{\partial\bar{\beta}_2}-(1\leftrightarrow 2)
\right|^{-1},
\label{final}
\eea
where $\Theta(x)$ is the Heaviside function. Due to the complicated
relation between energy/momentum and the spectral parameter $\beta$ we
have not been able to simplify \r{final} further. It is however in a
form that can be readily analyzed numerically. Let us first discuss 
constant momentum scans, i.e. the behviour of $S(\omega,p)$ as a function
of $\omega$ for fixed $p$. At the antiferromagnetic wave vector $p=\pi$ 
the singularity at zero frequency is
\bea
S(\omega,\pi)&\propto& \frac{1}{\omega}\sqrt{\log\frac{1}{\omega}}\ {\rm for\ 
\omega\to 0}\ ,\nn
S(\omega,\pi)&\propto& \sqrt{\omega_U(\pi)-\omega}\ {\rm for\ 
\omega\to \omega_U(\pi)}\ .
\eea
The power-law behaviour in the $\omega\to 0$ limit agrees with the result 
obtained from finite-size corrections and conformal field theory \cite{lowE}.
In Fig.2 we plot $S_2(\omega,\pi)$ on a logarithmic scale. One can see 
that the $\frac{1}{\omega}\sqrt{\log\frac{1}{\omega}}$ behaviour holds even 
for relatively large values of $\omega$.

\begin{figure}[ht]
\begin{center}
\noindent
\epsfxsize=0.45\textwidth
\epsfbox{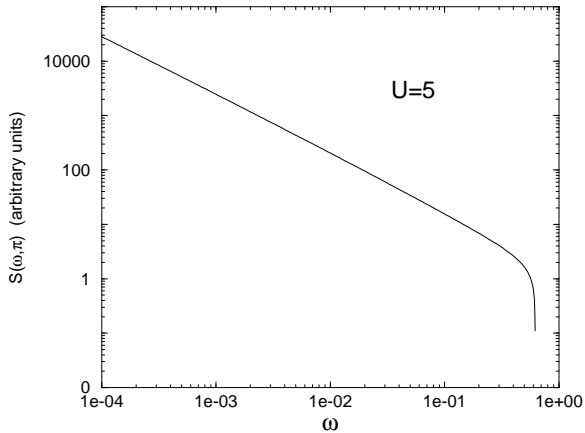}
\end{center}
\caption{\label{fig:cm}%
Constant momentum scan for $p=\pi$ and $U=5$
.} 
\end{figure}
In Fig.3 we plot $S_2(0.3,p)$ as a function of the momentum $p$. Obviously
the dynamical structure factor is nonzero only if intermediate states with
energy $\omega=0.3$ and momentum $p$ are available i.e.
inside the dispersion of the 2-spinon continuum shown in Fig.1. 

\begin{figure}[ht]
\begin{center}
\noindent
\epsfxsize=0.45\textwidth
\epsfbox{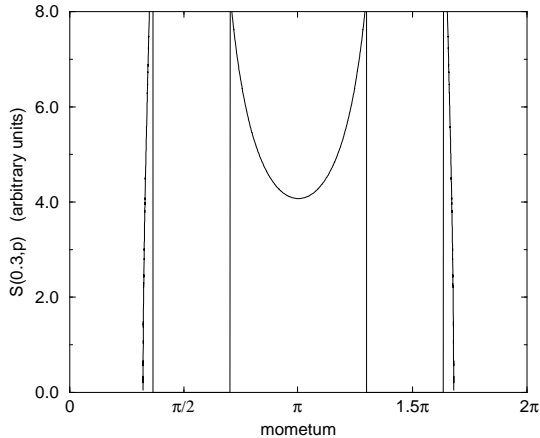}
\end{center}
\caption{\label{fig:ce}%
Constant energy scan for $\omega=0.3$ and $U=5$
.} 
\end{figure}

Finally, let us discuss consequences of our results for the attractive
Hubbard model. It is known that the half-filled repulsive Hubbard model
can be mapped to the half-filled attractive Hubbard model by means of the
unitary transformation defined by
\be
{\cal U} c_{j,\da} {\cal U}^\dagger = c_{j,\da} \ ,\qquad 
{\cal U} c_{j,\up} {\cal U}^\dagger = (-1)^j\ c^\dagger_{j,\up} \ ,
\ee
under which $H(U)$ goes to $H(-U)$ and the spin \r{spin} 
and eta-pairing operators \r{eta} get interchanged. The repulsive ground 
state is mapped into the attractive one \cite{lieb}, which in turn implies
that the two spinon states considered above get mapped into the
charge-wave states discussed in \cite{smat} (which have up to a shift in
momentum by $\pi$ the same dispersion as the two spinon states). 
As a consequence we can obtain
from \r{a-} the two charge-wave contribution to pairing and density-density 
correlation functions in the attractive half-filled Hubbard model
\bea
&&\langle 0| c^\dagger_{m,\up}(t)c^\dagger_{m,\da}(t)\ 
c_{0,\da}(0)c_{0,\up}(0)|0\rangle\ ,\nn
&&\langle 0| n_{m}(t) n_{0}(0)|0\rangle\ ,
\eea
where $n_m=c^\dagger_{m,\up}c_{m,\up}+c^\dagger_{m,\da}c_{m,\da}$. 
The analysis of the corresponding Fourier transforms is identical to
the one carried out above for the dynamical structure factor.

\section{Conclusions}

We have proposed an exact expression for the two spinon formfactor of 
spin operators in the half-filled Hubbard chain. We explicitly calculated
the corresponding contribution to the dynamical structure factor. Our
proposal \r{a-} can be extended to multispinon formfactors of spin 
operators \cite{unpub}. We think that very recent developments \cite{kmt} 
in the calculation of formfactors in integrable lattice models will
make it possible to prove \r{a-} rigourously and also lead to the
determination of formfactors of the fermion operators in the near future.

\vskip .5cm
\centerline{\bf Acknowledgements:}
\vskip .5cm
We thank the I.S.I. Torino, where part of this work was done, 
for hospitality. We are grateful to S. Pakuliak for correspondence and 
R.I. Nepomechie for discussions.
This work was supported by the NSF under grant number PHY-9605226 (V.E.K.) 
and by the EPSRC (F.H.L.E.).

\end{narrowtext}
\end{document}